# Description of electromagnetic fields in inhomogeneous accelerating sections. IV Couplers

M.I. Ayzatsky[1]

National Science Center "Kharkiv Institute of Physics and Technology" (NSC KIPT), 61108, Kharkiv, Ukraine

A new approach to incorporating coupling elements into a generalized coupled mode theory is presented. The simplest model of coupling of a structured waveguide with an external RF power source and load through loops and transmission lines was used. Even such a simple model significantly complicated the system of coupled equations – it turned into a coupled integro-differential system of the Barbashin type with degenerate kernels. Since the integral kernels are degenerate, this system is reduced to three independent systems of differential equations. Instead of solving a system of coupled integro-differential equations, we need to find solutions to three systems of ordinary differential equations. Two systems describe the distribution of the field excited by one loop and the specified value of the excitation current in it. In the first system the loop is located at the section's input, and in the second, at the section's output. The third system does not depend on the loop parameters at all. It describes the distribution of the field excited by an electron beam in a section without loops. Based on the developed analytical model, the computer code was developed for matching the loop couplers for the uniform accelerating sections of X-band. The calculation results were used to simulate the non-uniform section. Without additional matching, we obtained an input reflection coefficient of 8E-3.

## 1. Introduction

Waveguides that consist of similar (but not always identical) cells are called structured. Structured waveguides based on coupled resonators play an important role in many applications. Most of the travelling wave accelerating sections are belong to them. They are always inhomogeneous in the electrodynamic sense. Two types of inhomogeneity can exist. The first is associated with possible inhomogeneity of the internal cells. The second type of inhomogeneity, which is always present, is associated with couplers. Typically, these are the end cells connected to the waveguides. Due to the need to ensure full transmission, their dimensions differ significantly from those of the internal cells (see, for example, [1] and the literature cited there).

Recently it was proposed to use a modified uniform basis for description of non-periodic structured waveguides and a new generalization of the theory of coupled modes was constructed [2,3,4,5,6]. One of the positive features of proposed coupled modes theory is the simple and clear procedure for taking into account the beam loading and the easy transition to the case of a homogeneous waveguide [7,8]. It was shown that an infinitive system of equations can be reduced and the single mode representation describes the electromagnetic fields with small errors in the accelerating section [3,4]. Within the single-mode approximation, the fields are represented as the sum of two components, one of which is associated with the right-travelling eigenwave and the other with the left-travelling eigenwave.

In this paper, we propose a new approach to incorporating coupling elements into the generalized coupled mode theory. We used the simplest model of coupling of a structured waveguide with an external RF power source and load through loops and transmission lines. The results of calculation of the distribution of electric fields excited by an external RF source are presented.

## 2. Main equations

We will assume that the dependence of all quantities on time is $\exp(-i\omega t)$.

Electromagnetic fields in a non-periodic structured waveguide with the ideal metal walls can be represented in the form of such series ([2])

---

[1] E-mail: mykola.aizatsky@gmail.com, aizatsky@kipt.kharkov.ua

$$\vec{H}(\vec{r}) = \sum_{s=-\infty}^{s=\infty} C_s(z) \vec{H}_s^{(e,z)}(\vec{r}),$$

$$\vec{E}(\vec{r}) = \sum_{s=-\infty}^{s=\infty} C_s(z) \vec{E}_s^{(e,z)}(\vec{r}) + \frac{\vec{j}}{i\omega\varepsilon_0\varepsilon}, \quad (1)$$

where $\vec{E}_s^{(e,z)}(\vec{r}), \vec{H}_s^{(e,z)}(\vec{r})$ are modified eigen vector functions obtained by generalizing the eigen $\vec{E}_s^{(e)}, \vec{H}_s^{(e)}$ vectors of a homogeneous waveguide by special continuation of the geometric parameters ([2,3]). The eigen waves of a homogeneous waveguide we present as $(\vec{E}_s, \vec{H}_s) = (\vec{E}_s^{(e)}, \vec{H}_s^{(e)})\exp(\gamma_s z)$, where ($\vec{E}_s^{(e)}, \vec{H}_s^{(e)}$) are the periodic functions of the z-coordinate. Under such choice of the basis functions, the coefficients $C_s(z)$ include an exponential dependence on the z-coordinate.

Electromagnetic fields (1) must obey the Maxwell's equations. To satisfy this condition $C_s(z)$ must be solutions of such a coupled system of differential equations [3,4]

$$\frac{dC_s}{dz} - \gamma_s^{(e,z)} C_s + \frac{1}{2N_s^{(e,z)}} \frac{dN_s^{(e,z)}}{dz} C_s + \sum_{s'=-\infty}^{\infty} C_{s'} U_{s',s}^{(z)} = \frac{1}{N_s^{(e,z)}} \int_{S_\perp^{(z)}} \vec{j}_\omega \vec{E}_{-s}^{(e,z)} dS, \quad (2)$$

where

$$U_{s',s} = \frac{1}{2N_s^{(e,z)}} \sum_i \frac{dg_i^{(z)}}{dz} \int_{S_\perp^{(z)}(z)} \left\{ \left[\frac{\partial \vec{E}_{-s}^{(e,z)}}{\partial g_i^{(z)}} \vec{H}_{s'}^{(e,z)}\right] - \left[\vec{E}_{-s}^{(e,z)} \frac{\partial \vec{H}_{s'}^{(e,z)}}{\partial g_i^{(z)}}\right] + \left[\frac{\partial \vec{E}_{s'}^{(e,z)}}{\partial g_i^{(z)}} \vec{H}_{-s}^{(e,z)}\right] - \left[\vec{E}_{s'}^{(e,z)} \frac{\partial \vec{H}_{-s}^{(e,z)}}{\partial g_i^{(z)}}\right] \right\} \vec{e}_z dS, \quad (3)$$

$g_i^{(z)}(z)$ -generalized geometrical parameters, $\gamma_s^{(e,z)}(z)$ - generalized wavenumber and

$$N_s^{(e,z)} = \int_{S_\perp^{(z)}(z)} \left\{ \left[\vec{E}_s^{(e,z)} \vec{H}_{-s}^{(e,z)}\right] - \left[\vec{E}_{-s}^{(e,z)} \vec{H}_s^{(e,z)}\right] \right\} \vec{e}_z dS = \begin{cases} N_s^{(z)}, & s > 0, \\ -N_s^{(z)}, & s < 0. \end{cases} \quad (4)$$

We will consider axisymmetric TH (E) electromagnetic fields. In this case $\vec{E}_s^{(e,z)} = \vec{e}_r E_{r,s}^{(e,z)} + \vec{e}_z E_{z,s}^{(e,z)}$ and $\vec{H}_s^{(e,z)} = \vec{e}_\varphi H_{\varphi,s}^{(e,z)}$.

In the single mode approach ($s=1$), when the representation (1) transforms into

$$\vec{E}_1(\vec{r}) = \vec{E}_1^+(\vec{r}) + \vec{E}_1^-(\vec{r}) = C_1(z)\vec{E}_1^{(e,z)}(\vec{r}) + C_{-1}(z)\vec{E}_{-1}^{(e,z)}(\vec{r}), \quad (5)$$

the coupled system (2) is written as

$$\frac{dC_1}{dz} - \gamma_1^{(e,z)} C_1 + \frac{1}{2N_s^{(e,z)}} \frac{dN_s^{(e,z)}}{dz} C_1 + C_1 U_{1,1} + C_{-1} U_{-1,1} = \frac{1}{N_1^{(e,z)}} \int_{S_\perp^{(z)}} \vec{j}_\omega \vec{E}_{-1}^{(e,z)} dS$$

$$\frac{dC_{-1}}{dz} + \gamma_1^{(e,z)} C_{-1} + \frac{1}{2N_s^{(e,z)}} \frac{dN_s^{(e,z)}}{dz} C_2 + C_{-1} U_{-1,-1} + C_1 U_{1,-1} = -\frac{1}{N_1^{(e,z)}} \int_{S_\perp^{(z)}} \vec{j}_\omega \vec{E}_1^{(e,z)} dS \quad (6)$$

We will consider $\vec{E}_k^{(e,z)}$ as dimensionless vector, the vector $\vec{H}_k^{(e,z)}$ has a dimension $[\vec{H}_k^{(e,z)}] = \Omega^{-1}$.

Taking into account that $U_{-1,-1} = -U_{1,1}$ and introducing new functions $\tilde{C}_{\pm 1}$,

$$C_{\pm 1} = E_D \sqrt{\frac{N_1^{(e,z)}(0)}{N_1^{(e,z)}(z)}} \Gamma^{(\pm)}(z) \tilde{C}_{\pm 1}, \quad (7)$$

where

$$\Gamma^{(\pm)}(z) = \exp\left(\pm \int_0^z \left(\gamma_1^{(e,z)} - U_{1,1}\right) dz'\right), \quad (8)$$

the system (6) takes the form

$$\frac{d\tilde{C}_1}{dz} = -\Gamma^{(-)2} U_{-1,1}\tilde{C}_{-1} + \frac{\Gamma^{(-)}}{E_D \sqrt{N_1^{(e,z)}(0) N_1^{(e,z)}(z)}} \int_{S_\perp^{(z)}} \vec{j}_\omega \vec{E}_{-1}^{(e,z)} dS,$$
$$\frac{d\tilde{C}_{-1}}{dz} = -\Gamma^{(+)2} U_{1,-1}\tilde{C}_1 - \frac{\Gamma^{(+)}}{E_D \sqrt{N_1^{(e,z)}(0) N_1^{(e,z)}(z)}} \int_{S_\perp^{(z)}} \vec{j}_\omega \vec{E}_1^{(e,z)} dS,$$
(9)

$E_D$ is a coefficient which has the dimension $[E_D]$=V/m. In the following we take $E_D$=1 MV/m and the values of $C_{\pm 1}$ and $\vec{E}_1^\pm$ in all figures are given in MV/m.

To solve the system of differential equations (9) we need to know the characteristics of eigen waves $\vec{E}_{\pm 1}^{(e,z)}(z), N_1^{(e,z)}(z), \gamma_1^{(e,z)}(z)$ as functions of the longitudinal coordinate $z$ and their derivatives $U_{1,1}(z), U_{-1,1}(z), U_{1,-1}(z)$.

We will consider the segmented waveguides which is formed from cylindrical regions (see Figure 1). For such geometry there is analytical-numerical approach which is based on the theory of coupled integral equations [9,10]. Using this approach, we developed a method of calculating the required characteristics [3,4]. In this method the fields in each cross section are represented as

$$E_r^{(k,q)} = \sum_m \bar{E}_r^{(k,q)}(\tilde{z}) J_1\left(\frac{\lambda_m}{b_{k,q}}r\right) = \sum_m \left\{ B_{m,1}^{(k,q)} \exp\left(\chi_m^{(k,q)}\tilde{z}\right) + B_{m,2}^{(k,q)} \exp\left(-\chi_m^{(k,q)}\tilde{z}\right) \right\} J_1\left(\frac{\lambda_m}{b_{k,q}}r\right),$$
(10)

$$E_z^{(k,q)} = \sum_m \bar{E}_z^{(k,q)}(\tilde{z}) J_0\left(\frac{\lambda_m}{b_{k,q}}r\right) = -\sum_m \frac{\lambda_m}{\chi_m^{(k,q)} b_{k,q}} \left\{ B_{m,1}^{(k,q)} \exp\left(\chi_m^{(k,q)}\tilde{z}\right) - B_{m,2}^{(k,q)} \exp\left(-\chi_m^{(k,q)}\tilde{z}\right) \right\} J_0\left(\frac{\lambda_m}{b_{k,q}}r\right),$$
(11)

$$H_\varphi^{(k,q)} = \sum_m \bar{H}_\varphi^{(k,q)}(\tilde{z}) J_1\left(\frac{\lambda_m}{b_{k,q}}r\right) = i\frac{\omega b_{k,q}}{c}\frac{1}{Z_0}\sum_m \frac{\varepsilon}{\chi_m^{(k,q)} b_{k,q}} \left\{ B_{m,1}^{(k,q)} \exp\left(\chi_m^{(k,q)}\tilde{z}\right) - B_{m,2}^{(k,q)} \exp\left(-\chi_m^{(k,q)}\tilde{z}\right) \right\} J_1\left(\frac{\lambda_m}{b_{k,q}}r\right),$$
(12)

where $Z_0$ is vacuum impedance, $B_{m,1}^{(k,q)}, B_{m,2}^{(k,q)}, \chi_m^{(k,q)}, b_{k,q}$ are the functions of $z$ (see [2,3]). The use of such a representation allows us to significantly simplify the numerical calculation of magnetic flux (see (16)).

The current $\vec{j}_\omega$ is the sum of the beam current and the current in the loop
$$\vec{j}_\omega = \vec{j}_{beam} + \vec{j}_{loop}.$$
(13)

Assuming that the current along the conductor is constant, the current in the loop can be written as
$$\vec{j}_{loop} = \vec{G}(\vec{r}) i(0,t) = \vec{G}(\vec{r}) \left\{ \frac{2\tilde{u}^{(+)}(0)}{Z_v} - \frac{1}{Z_v}\frac{d\Phi}{dt} \right\} = \vec{G}(\vec{r}) \left\{ \frac{2\tilde{u}^{(+)}(0)}{Z_v} + i\omega \frac{\Phi}{Z_v} \right\},$$
(14)

where $Z_v$ is the characteristic impedance of the transmission line, $\tilde{u}^{(+)}(\xi)$ is the Fourier amplitude of the wave propagating along the transmission line in the direction of the coupler loop ($u^+(t,\xi) = \tilde{u}^{(+)}(\xi)\exp(-i\omega t) + c.c.$), $\Phi = \int_{S_g} \vec{B} d\vec{S}$ is the magnetic field flux through the loop. The geometric factor $\vec{G}(\vec{r})$ for a thin conductor is

$$\vec{G}(\vec{r}) = \delta(\varphi)\frac{1}{r}\delta(r - r_1 + r_g(z))\vec{e}_g(z), \quad z_1 < z < z_2,$$
(15)

where the function $r_g(z)$ is the distance from the loop point with the longitudinal coordinate $z$ to the side wall of the resonator (see Figure 1), $\vec{e}_g(z)$ is a unit vector tangent to loop at the same point.

The magnetic field flux can be reduced to a sum of one-dimensional integrals

$$\Phi = \int_{S_g} \vec{B} d\vec{S} = \mu_0 \int_{S_g} \left\{ C_{+1}(z)\vec{H}_{+1}^{(e,z)}(\vec{r}) + C_{-1}(z)\vec{H}_{-1}^{(e,z)}(\vec{r}) \right\} d\vec{S} = \frac{1}{i\omega}\int_{z_1}^{z_2} dz \left\{ \sigma_{+1}(z)C_{+1}(z) + \sigma_{-1}(z)C_{-1}(z) \right\},$$
(16)

where $\sigma_{\pm 1}(z) = \frac{i\omega}{c}\int_{R_1 - r_g(z)}^{R_1} dr H_{\pm 1,\varphi}^{(e,z)}(r,z) = b_{1,2}\frac{1}{Z_0}\frac{i\omega}{c}\sum_m \bar{H}_{\varphi,m}^{(1,2,\pm)}\frac{1}{\lambda_m}J_0\left(\frac{\lambda_m}{b_{1,2}}(b_{1,2} - r_g(z))\right)$, $\sigma_{\pm 1}(z_1) = \sigma_{\pm 1}(z_2) = 0.$.

The integrals in equations (9) we transform into

$$\int_{S_\perp^{(z)}} \vec{j}_\omega \vec{E}_{\pm 1}^{(e,z)} dS = \int_{S_\perp^{(z)}} \vec{j}_{beam} \vec{E}_{\pm 1}^{(e,z)} dS + \int_{S_\perp^{(z)}} \vec{j}_{loop} \vec{E}_{\pm 1}^{(e,z)} dS = \int_{S_\perp^{(z)}} \vec{j}_{beam} \vec{E}_{\pm 1}^{(e,z)} dS + \Upsilon_{\pm 1}(z) \left\{ \frac{2\tilde{u}^{(+)}(0)}{Z_v} + i\omega \frac{\Phi}{Z_v} \right\} = \\ = \int_{S_\perp^{(z)}} \vec{j}_{beam} \vec{E}_{\pm 1}^{(e,z)} dS + \frac{\Upsilon_{\pm 1}(z)}{Z_v} \left[ 2\tilde{u}^{(+)}(0) + \int_{z_1}^{z_2} dz' \{ \sigma_{+1}(z') C_{+1}(z') + \sigma_{-1}(z') C_{-1}(z') \} \right], \quad (17)$$

where

$$\Upsilon_{\pm 1}(z) = \begin{cases} 0, \ z < z_1 \\ -E_{\mp,r}^{(e,z)}(r_1 - r_g(z), z) \sin \varphi_g(z) + E_{\mp,z}^{(e,z)}(r_1 - r_g(z), z) \cos \varphi_g(z), \ z_1 < z < z_2 \\ 0, \ z > z_2 \end{cases} \quad (18)$$

$$tg\, \varphi_g(z) = \frac{dr_g(z)}{dz}.$$

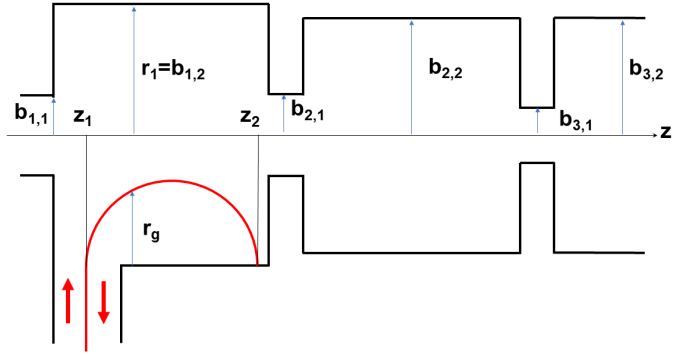

*Figure 1 Schematic representation of the initial part of the section under consideration*

The travelling wave section usually has two couplers for supplying high-frequency energy to the section and to remove unused portions of energy from it. In this case, the final system of equations will look like this

$$\frac{d\tilde{C}_1}{dz} = \mu_1(z)\tilde{C}_{-1} + \tilde{\Upsilon}_{1,1}(z)\Xi_1 + \tilde{\Upsilon}_{1,2}(z)\Xi_2 + \tilde{I}_1(z), \\ \frac{d\tilde{C}_{-1}}{dz} = \mu_{-1}(z)\tilde{C}_1 - \tilde{\Upsilon}_{-1,1}(z)\Xi_1 - \tilde{\Upsilon}_{-1,2}(z)\Xi_2 - \tilde{I}_{-1}(z). \quad (19)$$

where the second subscript index $i$ in $\tilde{\Upsilon}_{\pm 1,i}(z)$ refers to the input coupler ($i=1$, $z_1 < z < z_2$) and to the output one ($i=2$, $z_3 < z < z_4$) and

$$\Xi_1 = \int_{z_1}^{z_2} dz' \left( \tilde{\Gamma}_{1,1}(z')\tilde{C}_1(z') + \tilde{\Gamma}_{-1,1}(z')\tilde{C}_{-1}(z') \right) + 2\frac{\tilde{u}^{(+)}}{E_D} \\ \Xi_2 = \int_{z_3}^{z_4} dz' \left( \tilde{\Gamma}_{1,2}(z')\tilde{C}_1(z') + \tilde{\Gamma}_{-1,2}(z')\tilde{C}_{-1}(z') \right) \quad (20)$$

$$\mu_{\pm 1}(z) = -\Gamma_{\mp 1}^2(z) U_{\mp 1,\pm 1}(z),$$

$$\tilde{\Upsilon}_{\pm 1,i}(z) = \frac{1}{\sqrt{\tilde{N}_1^{(e,z)}(0)\tilde{N}_1^{(e,z)}(z)}} \frac{\Upsilon_{\pm 1,i}(z)\Gamma_{\mp 1}(z)Z_0}{Z_v},$$

$$\tilde{\Gamma}_{\pm 1,i}(z) = \sqrt{\frac{\tilde{N}_1^{(e,z)}(0)}{\tilde{N}_1^{(e,z)}(z)}} \Gamma_{\pm 1}(z)\sigma_{\pm 1,i}(z), \qquad (21)$$

$$\tilde{I}_{\pm 1}(z) = \frac{\Gamma_{\mp 1} Z_0}{E_D \sqrt{\tilde{N}_1^{(e,z)}(0)\tilde{N}_1^{(e,z)}}} \int_{S_\perp^{(z)}} \vec{j}\vec{E}_{\mp 1}^{(e,z)} dS,$$

$$\tilde{N}_1^{(e,z)}(z) = N_1^{(e,z)}(z)Z_0.$$

We need to add the boundary conditions for the functions $\tilde{C}_{\pm 1}$. Along the central axis, the travelling wave accelerating section begins and ends with the "cut-off" waveguides. In our consideration we will suppose that these "cut-off" waveguides end with the metal walls. For such configuration the boundary conditions for the functions $\tilde{C}_{\pm 1}$ are written in the following form:

$$\tilde{C}_{-1}(0) = \tilde{C}_{+1}(0)\eta_1, \quad \tilde{C}_{-1}(L) = \tilde{C}_{+1}(L)\eta_2,$$
$$\eta_1 = \Gamma^{(+)2}(0) E_{r,+1}^{(e,z)}(0) / E_{r,-1}^{(e,z)}(0), \quad \eta_2 = \Gamma^{(+)2}(L) E_{r,+1}^{(e,z)}(L) / E_{r,-1}^{(e,z)}(L). \qquad (22)$$

System of linear integro-differential equations (19) is of Barbashin type (see, for example,[11]) and has the degenerate kernels. In this case we can seek the solutions of the system (19) in the form

$$\tilde{C}_1 = \Xi_1 \tilde{C}_{1,1} + \Xi_2 \tilde{C}_{1,2} + \tilde{C}_{1,3},$$
$$\tilde{C}_{-1} = \Xi_1 \tilde{C}_{-1,1} + \Xi_2 \tilde{C}_{-1,2} + \tilde{C}_{-1,3}, \qquad (23)$$

where $\tilde{C}_{\pm 1,1}, \tilde{C}_{\pm 1,2}, \tilde{C}_{\pm 1,3}$, are six new unknown functions.

Substituting (23) into (19) gives

$$\Xi_1\left(\frac{d\tilde{C}_{1,1}}{dz} - \mu_1 \tilde{C}_{-1,1} - \tilde{\Upsilon}_{1,1}(z)\right) + \Xi_2\left(\frac{d\tilde{C}_{1,2}}{dz} - \mu_1 \tilde{C}_{-1,2} - \tilde{\Upsilon}_{1,2}(z)\right) + \left(\frac{d\tilde{C}_{1,3}}{dz} - \mu_1 \tilde{C}_{-1,3} - \tilde{I}_1\right) = 0$$
$$\Xi_1\left(\frac{d\tilde{C}_{-1,1}}{dz} - \mu_{-1} \tilde{C}_{1,1} + \tilde{\Upsilon}_{-1,1}(z)\right) + \Xi_2\left(\frac{d\tilde{C}_{-1,2}}{dz} - \mu_{-1} \tilde{C}_{1,2} + \tilde{\Upsilon}_{-1,2}(z)\right) + \left(\frac{d\tilde{C}_{-1,3}}{dz} - \mu_{-1} \tilde{C}_{1,3} + \tilde{I}_{-1}\right) = 0 \qquad (24)$$

Instead of two unknown functions $\tilde{C}_{\pm 1}$ we introduced six new unknown functions $\tilde{C}_{\pm 1,1}, \tilde{C}_{\pm 1,2}, \tilde{C}_{\pm 1,3}$, , so we can impose four additional conditions, which we write as follows:

$$\left.\begin{array}{l}\dfrac{d\tilde{C}_{1,1}}{dz} = \mu_1 \tilde{C}_{-1,1} + \tilde{\Upsilon}_{1,1}(z),\\[4pt] \dfrac{d\tilde{C}_{-1,1}}{dz} = \mu_{-1} \tilde{C}_{1,1} - \tilde{\Upsilon}_{-1,1}(z),\end{array}\right\} \tilde{C}_{-1,1}(0) = \eta_1 \tilde{C}_{1,1}(0), \tilde{C}_{-1,1}(L) = \eta_2 \tilde{C}_{1,1}(L),$$

$$\left.\begin{array}{l}\dfrac{d\tilde{C}_{1,2}}{dz} = \mu_1 \tilde{C}_{-1,2} + \tilde{\Upsilon}_{1,2}(z),\\[4pt] \dfrac{d\tilde{C}_{-1,2}}{dz} = \mu_{-1} \tilde{C}_{1,2} - \tilde{\Upsilon}_{-1,2}(z),\end{array}\right\} \tilde{C}_{-1,2}(0) = \eta_1 \tilde{C}_{1,2}(0), \tilde{C}_{-1,2}(L) = \eta_2 \tilde{C}_{1,2}(L). \qquad (25)$$

As $\Xi_1, \Xi_2 \neq 0$, then from (24) we obtain additional two equations

$$\left.\begin{array}{l}\dfrac{d\tilde{C}_{1,3}}{dz} = \mu_1 \tilde{C}_{-1,3} + \tilde{I}_1 = 0,\\[4pt] \dfrac{d\tilde{C}_{-1,3}}{dz} = \mu_{-1} \tilde{C}_{1,3} - \tilde{I}_{-1} = 0,\end{array}\right\} \tilde{C}_{-1,3}(0) = \eta_1 \tilde{C}_{1,3}(0), \tilde{C}_{-1,3}(L) = \eta_2 \tilde{C}_{1,3}(L). \qquad (26)$$

Substituting (23) into (20), we get the system of linear equations for $\Xi_1, \Xi_2$

$$\Xi_1(1-P_{1,1}) - \Xi_2 P_{1,2} = P_{1,3} + 2\frac{u_+}{E_D},$$
$$\Xi_1 P_{2,1} - \Xi_2(1-P_{2,2}) = -P_{2,3},$$
(27)

where

$$P_{1,i} = \int_{z_1}^{z_2} dz'\left(\tilde{\Gamma}_1(z')\tilde{C}_{1,i}(z') + \tilde{\Gamma}_{-1}(z')\tilde{C}_{-1,i}(z')\right)$$

$$P_{2,i} = \int_{z_3}^{z_4} dz'\left(\tilde{\Gamma}_1(z')\tilde{C}_{1,i}(z') + \tilde{\Gamma}_{-1}(z')\tilde{C}_{-1,i}(z')\right)$$
(28)

Solution of this system is

$$\Xi_1 = \frac{\left(P_{1,3} + 2\frac{u_+}{E_D}\right)(1-P_{2,2}) + P_{2,3}P_{1,2}}{(1-P_{1,1})(1-P_{2,2}) - P_{1,2}P_{2,1}},$$

$$\Xi_2 = \frac{P_{2,1}\left(P_{1,3} + 2\frac{u_+}{E_D}\right) + (1-P_{1,1})P_{2,3}}{(1-P_{1,1})(1-P_{2,2}) - P_{1,2}P_{2,1}}.$$
(29)

Amplitudes of the waves propagating in the transmission lines in the direction from the couplers ($u^-(t,\xi) = \tilde{u}^{(-)}(\xi)\exp(-i\omega t) + c.c.$) can be found from the following relationships

$$\tilde{u}_1^-(0) = E_D\left[-\int_{z_1}^{z_2} dz'\left(\tilde{\Gamma}_{1,1}(z')\tilde{C}_1(z') + \tilde{\Gamma}_{-1,1}(z')\tilde{C}_{-1}(z')\right) - \frac{\tilde{u}_1^{(+)}(0)}{E_D}\right] =$$
$$= -E_D\left[\Xi_1 P_{1,1} + \Xi_2 P_{1,2} + P_{1,3} + \frac{\tilde{u}_1^+(0)}{E_D}\right]$$
(30)

$$\tilde{u}_2^{(-)}(0) = -E_D\left[\Xi_1 P_{2,1} + \Xi_2 P_{2,2} + P_{2,3}\right]$$
(31)

So, instead of solving the system of coupled integro-differential equations (19) we need to find the solutions $\tilde{C}_{\pm 1,i}(z)$ to three systems of ordinary differential equations (25) - (26), calculate six integrals (28), find constants $\Xi_1, \Xi_2$. By solving these problems, we can find the amplitudes $C_1(z), C_{-1}(z)$ and the field distribution in the section $\vec{E}_1(\vec{r}) = \vec{E}_1^+(\vec{r}) + \vec{E}_1^-(\vec{r}) = C_1(z)\vec{E}_1^{(e,z)}(\vec{r}) + C_{-1}(z)\vec{E}_{-1}^{(e,z)}(\vec{r})$.

It's worth to note that the boundary problems (25) describe the distribution of the field excited by one loop and the specified value of the excitation current in it. In the first problem, the loop is located at the section's input, and in the second, at the section's output. What is unexpected is that the third boundary problem (26) does not depend on the loop parameters at all. It describes the distribution of the field excited by an electron beam in a section without loops.

For numerical analysis of the coupling systems (25)-(26) we used method based on the 4th order Runge-Kutta method [6,12,13].

The systems (25)-(26) have two peculiarities. First peculiarity is the presence of an integral $\Gamma^{(\pm)}(z)$ with a variable upper limit. To calculate it, we used Simpson formula. Second, since the thickness of the disks ($d_{1,k}$) and the length of the resonators ($d_{2,k}$) are not constant and can differ significantly we cannot use the mesh with constant step. We chose the mesh $\{z_n\}$, $n = 1 \div N$, where the number of divisions of each segment (disk, resonator) $N_D$ is constant. We usually took $N_D = 60$. With such discretization, the step is not constant along the structure. To implement the Runge-Kutta method for each segment with a step $h_k$ it is necessary to divide the interval [$z_n, z_{n+1} = z_n + h_k$] into four equal parts and, using the Simpson formular, calculate the values of the integral and right parts in two points: $z = z_n + (z_{n+1} - z_n)/2$ and $z = z_{n+1}$.

A semi-elliptical shape with semi-axes $g_r, g_z$ was chosen as the shape of the loop.

The resulting system of linear equations for any boundary problem (25)-(26) has the form (see [6])

$$\begin{aligned}
&\left(\alpha_{1,1}(1)+\alpha_{1,2}(1)\eta_1\right)X_{1,1} - X_{1,2} = -W_{1,1}, \\
&\left(\alpha_{2,1}(1)+\alpha_{2,2}(1)\eta_1\right)X_{1,1} - X_{2,2} = -W_{2,1}, \\
&\left.\begin{array}{l}\alpha_{1,1}(n)X_{1,n}+\alpha_{1,2}(n)X_{2,n}-X_{1,n+1} = -W_{1,n} \\ \alpha_{2,1}(n)X_{1,n}+\alpha_{2,2}(n)X_{2,n}-X_{2,n+1} = -W_{2,n}\end{array}\right\} n=2,\ldots,N-2, \\
&\alpha_{1,1}(N-1)X_{1,N-1}+\alpha_{1,2}(N-1)X_{2,N-1}-X_{1,N} = -W_{1,N-1}, \\
&\alpha_{2,1}(N-1)X_{1,N-1}+\alpha_{2,2}(N-1)X_{2,N-1}-\eta_2 X_{1,N} = -W_{2,N-1},
\end{aligned} \quad (32)$$

where $X_{1,i}$ and $X_{2,i}$ are any pair of sequences $\tilde{C}_{\pm 1,1}(z_i), \tilde{C}_{\pm 1,2}(z_i), \tilde{C}_{\pm 1,3}(z_i)$.

System of equations (32) is sparce, so to solve it we used a special numerical procedures LSLZG from the IMSL MATH LIBRARY.

### 3. Couplers for the section with homogeneous DLW

In our previous works on the theory of inhomogeneous accelerating sections [2-6] we were considering the section based on a disk-loaded waveguide (DLW), which is an analogue of the one developed at CERN [14]. This section is unique. Not only do the resonator diameters and hole sizes in the diaphragms change [2], but the diaphragm thicknesses also decrease along the structure[3]. The working frequency of this section is $f = 11.994$ GHz, phase advance per cell is 2π/3. We will continue to use this section as a prototype. Bellow we will assume that $\tilde{u}_1^+(0) = 10 kV$, $P_+ = 1.06$ MW.

If the cells of the TW section are designed such that the phase velocity of the operating mode is equal to the beam velocity, then the coupler cell dimensions are designed in order to create a travelling wave regime in a section (output coupler) and minimize the reflected power at the section input (input coupler).

The coupling matching procedure we used within the generalized coupled mode theory approach is as follows. In the approach described above, in the case of homogeneous DLW $U_{1,1}(z) = U_{1,-1}(z) = U_{-1,1}(z) = 0$, $\tilde{N}_1^{(e,z)}(z) = const$, $\gamma_1^{(e,z)}(z) = const$ and the two field components $\vec{E}_1^+(\vec{r})$ and $\vec{E}_1^-(\vec{r})$ represent the waves[4] propagating in forward and backward directions, respectively. Thus, to match the output coupler, we need to find conditions under which the amplitude of the left-propagating wave $\tilde{C}_{-1}$ is small in the homogeneous part of the section. In our calculations, we aimed to find the minimum value[5] of $R_{out} = |\tilde{C}_{-1}(z_*)/\tilde{C}_1(z_*)|$ by tuning the resonant frequency of the last cell (changing its radius $b_{2,N}$) and the sizes of the output loop. We examined the sections consisting of seven cells of a homogeneous DLW and two cells containing loops. We selected the coordinate $z_*$ in the middle of the fifth cell.

We accepted that the section begins and ends with the "cut-off" waveguides and we supposed that these "cut-off" waveguides end with the metal walls It is convenient to choose the length of the "cut-off" waveguides equal the half of the disk thickness as in this case without couplers we get a resonator with an eigen frequency equals the working one. In this case the region of inhomogeneity ($U_{i,j}(z) \neq 0$, $\tilde{N}_1^{(e,z)}(z) \neq const$,) will be localized in the diaphragms separating the coupler cells from the homogeneous part of the section where the transition from the coupler resonator to the first (last) resonator of the DLW occurs (for more details, see [2,3]). In the coupler cells eigenvectors $\vec{E}_1^{(e,z)}(\vec{r}), \vec{E}_{-1}^{(e,z)}(\vec{r})$ were calculated at a constant disk thickness equals to the thickness of the first diaphragm.

Dependencies of $U_{i,j}(z)$, $\tilde{N}_1^{(e,z)}(z)$, $\gamma_1^{(e,z)}(z)$ on $z$ are presented in Figure 2-Figure 5 for the matched geometry (the DLW with a hole radius of 3.15 mm and a disk thickness of 1.67 mm). It can be seen that the coupling

---

[2] Input, Output iris radii - 3.15÷2.35 mm

[3] Iris thickness - 1.67 ÷ 1.00 mm, structure period D=8.3317 mm **Error! Bookmark not defined.**.

[4] Sometimes they are called forward and backward waves. But these terms are more often used for waves with coinciding and opposite directions of phase and group velocities.

[5] Usually, we were able to get $|\tilde{C}_{-1}/\tilde{C}_1| \leq 10^{-3}$, but the question usually arises about the accuracy of calculating such small values

coefficients $U_{-1,1}$ and $U_{1,-1}$ are significant in the region of the first disk. The shift of the local logitudinal wavenumber $U_{1,1}$ changes with the change of sign, so its influence is small. The value that undergoes the greatest changes on the disk is the local logitudinal wavenumber $\gamma_1^{(e,z)}$ (Figure 5). This is determined by the large difference between the radii of the coupler cell and the first DLW cell.

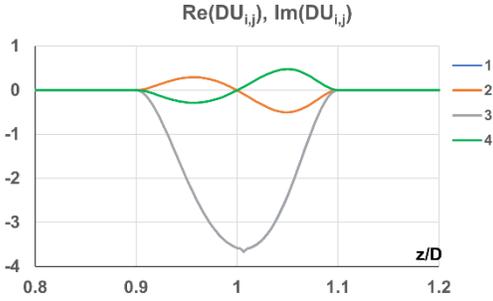

Figure 2 Dependences on $z$ of the coupling coefficients $U_{i,j}$ near the first disk: 1 -Re($DU_{-1,1}$), 2 -Im ($DU_{-1,1}$), 3- Re($DU_{1,-1}$) 4 - Im($DU_{1,-1}$)

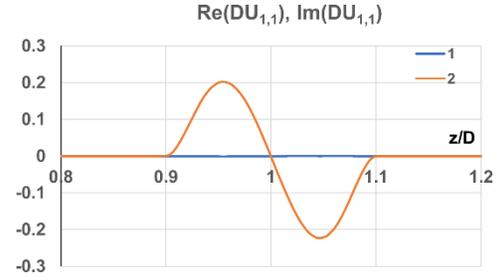

Figure 3 Dependences on $z$ of the shift of the local logitudinal wavenumber near the first disk: 1-Re($DU_{1,1}$), 2- Im($DU_{1,1}$),

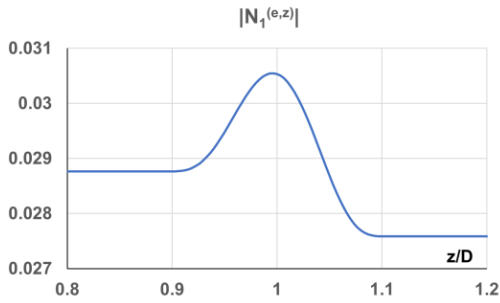

Figure 4 Dependences of the modulus of the norm $\tilde{N}_1^{(e,z)}$ on $z$ near the first disk

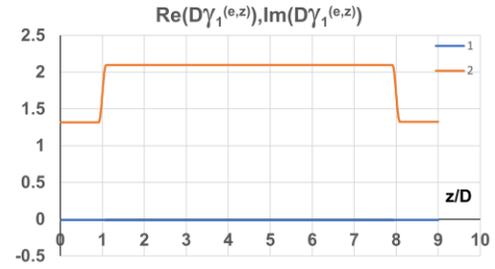

Figure 5 Dependences of the local logitudinal wavenumber $\gamma_1^{(e,z)}$ on $z$ : 1-Re($D\gamma_1^{(e,z)}$), 2- Im($D\gamma_1^{(e,z)}$)

Matching the input coupler was conducted by minimizing the reflection coefficient from the input coupler $R_{in} = \left| \tilde{u}_1^- / \tilde{u}_1^+ \right|$.

Results of matching the couplers for the DLW with a hole radius of 3.15 mm and a disk thickness of 1.67 mm are shown in Figure 6. Frequency dependence of reflection coefficient modules $R_{in}, R_{out}$ are shown on Figure 7.

The section without loops is a resonator and it has its own eigen frequencies. For the section under consideration, the eigen frequencies closest to the operating frequency[6] $f_0 = 11.994$ GHz are $f_L = 11.984$ GHz and $f_R = 12.010$ GHz. The solutions to systems (25)-(26) $\tilde{C}_{\pm 1,i}(z)$ depend on the ratio of the eigen frequencies to the frequency of an exciting force and are resonant. However, the amplitudes $C_1(z), C_{-1}(z)$ which are the sums of particular solutions (see (23)) have different resonant frequencies. At these frequencies the reflection coefficient $R_{in} = \left| \tilde{u}_1^- / \tilde{u}_1^+ \right|$ has minimum values.

---

[6] The eigen frequency of unmatched section equals $f_0$

From the spatial distributions of the characteristics of the amplitudes $C_1(z), C_{-1}(z)$ (see Figure 8 and Figure 9) it follows that the amplitudes $C_1(z)$ describes the right-travelling wave and the amplitudes $C_{-1}(z)$ describes a wave propagating to the left, since the its phase decreases with the coordinate $z$.

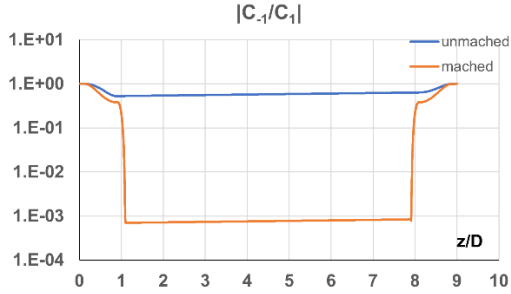

Figure 6 Spatial distributions of the modulus of the ratio of the amplitudes $|\tilde{C}_{-1}(z)/\tilde{C}_1(z)|$

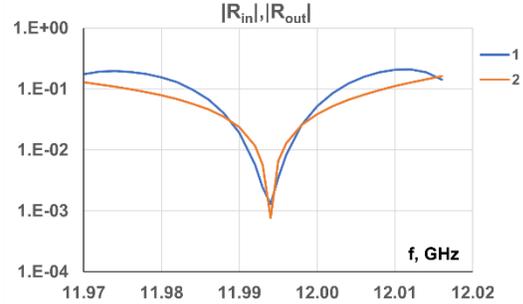

Figure 7 Frequency dependence of the modulus of the reflection coefficients: 1- $R_{in}$, 2- $R_{out}$

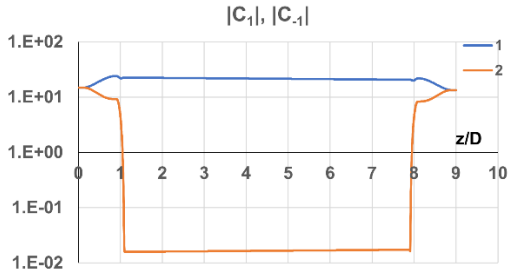

Figure 8 Spatial distributions of the moduli of the amplitudes: 1 - $\tilde{C}_1(z)$, 2 - $\tilde{C}_{-1}(z)$

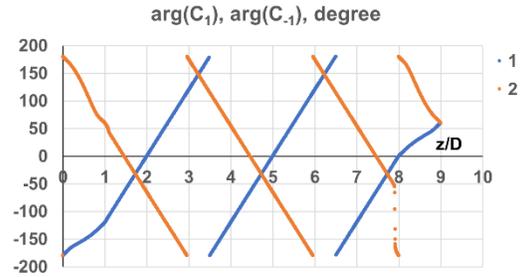

Figure 9 Spatial distributions of the phase of the amplitudes: 1 - $\arg\tilde{C}_1(z)$, 2 - $\arg\tilde{C}_{-1}(z)$

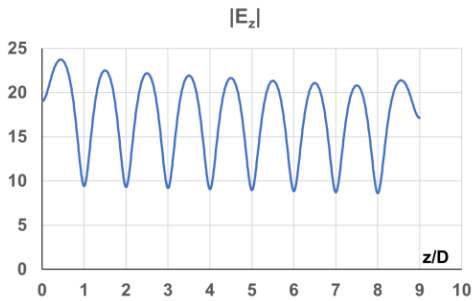

Figure 10 Spatial distributions of the modulus of the longitudinal electric field $E_z(r,z)|_{r=0}$,

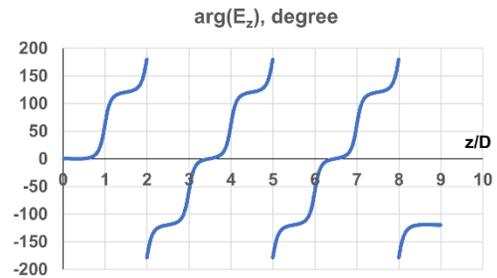

Figure 11 Spatial distributions of the phase of the longitudinal electric field $E_z(r,z)|_{r=0}$

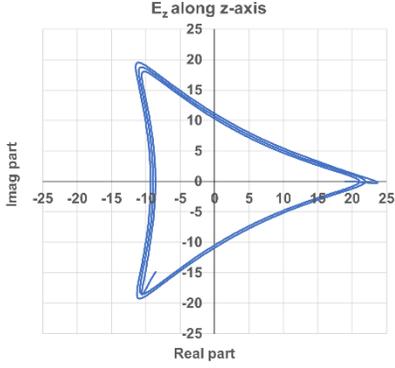

Figure 12 Trajectory in the complex plane of the longitudinal electric field $E_z(r,z)\big|_{r=0}$,

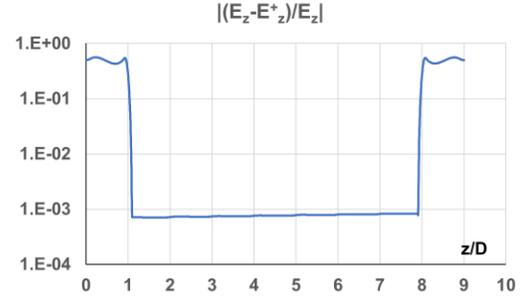

Figure 13 Modulus of relative error in the representation of the longitudinal electric field at $r=0$:

The spatial distributions of the characteristics of the longitudinal component of the full electric field $E_{1,z}(\vec{r}) = E_{1,z}^{+}(0,z) + E_{1,z}^{-}(0,z) = C_1(z)E_{1,z}^{(e,z)}(0,z) + C_{-1}(z)E_{-1,z}^{(e,z)}(0,z)$ on the axis $z$ are presented in Figure 10-Figure 12. We see that matching of the couplers ensured the traveling wave mode in the homogeneous part of the section. In the coupler regions there are the standing wave mode which is confirmed by the almost complete absence of phase dependence on the coordinate in the first and last cells. This circumstance proves the necessity of using the second component of electric field $\vec{E}_1^{-}(\vec{r})$ for correct describing of fields in an inhomogeneous waveguide. Figure 13 shows the error in calculating the field distribution using only the $\vec{E}_1^{+}(\vec{r})$ component, which is connected with the right travelling wave. The error in the coupler areas is very large.

### 4. Section with inhomogeneous DLW

Accelerating section CERN [14] has the first cell with a hole radius of 3.15 mm and a disk thickness of 1.67 mm and the last cell with a hole radius of 2.35 mm and a disk thickness of 1mm. We matched the input and output couplers that correspond to the homogeneous DLWs with such sizes and connected them with the nonuniform DLW which was designed earlier on the base of the Coupled Integral Equations Method [9]. Without additional matching we get the input reflection coefficient equals 8E-3.

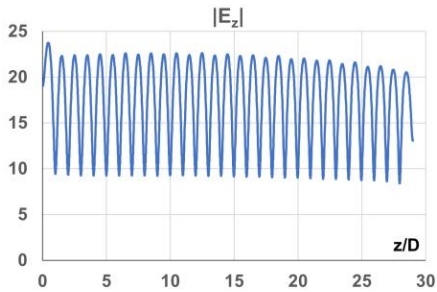

Figure 14 Spatial distributions of the modulus of the longitudinal electric field $E_z(r,z)\big|_{r=0}$

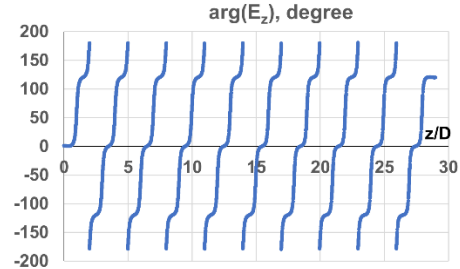

Figure 15 Spatial distributions of the phase of the longitudinal electric field $E_z(r,z)\big|_{r=0}$

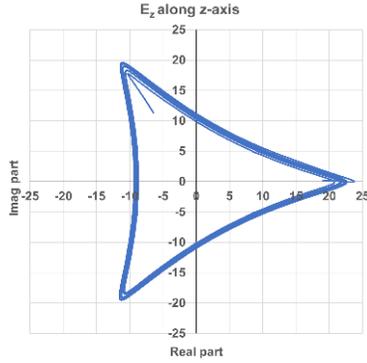

Figure 16 Trajectory in the complex plane of the longitudinal electric field $E_z(r,z)|_{r=0}$ ,

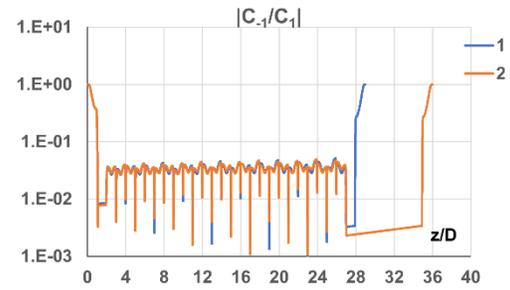

Figure 17 Spatial distributions of the modulus of the ratio of the amplitudes $|\tilde{C}_{-1}(z)/\tilde{C}_1(z)|$ : 1- section with 29 cells, 2 - section with additional identical cells before the output coupler

The longitudinal electric field $E_z(r,z)|_{r=0}$ , on the axis $z$ has the required spatial distribution (see Figure 14-Figure 16). For the structure with the homogeneous DLW the two field components $\vec{E}_1^+(\vec{r})$ and $\vec{E}_1^-(\vec{r})$ represent the waves propagating in forward and backward directions, respectively. In the case of the structure with the inhomogeneous DLW the field component $\vec{E}_1^-(\vec{r})$ transforms into the wave propagating in the forward direction and has a complicated spatial distribution (see [3-6]). The spatial distributions of the modulus of the ratio of the amplitudes $|\tilde{C}_{-1}(z)/\tilde{C}_1(z)|$ shown in Figure 17 differ significantly from the distribution shown in Figure 6. To show that this behavior is not due to reflection from the output coupler, we presented the same characteristic for a section with additional identical cells before the output coupler (see Figure 17).

### 4. Conclusions

A new approach to incorporating coupling elements into a generalized coupled mode theory is presented in this work. The simplest model of coupling of a structured waveguide with an external RF power source and load through loops and transmission lines was used. Even such a simple model significantly complicated the system of coupled equations – it turned into a coupled integro-differential system of the Barbashin type with degenerate kernels. Since the integral kernels are degenerate, this system is reduced to three independent systems of differential equations. Instead of solving a system of coupled integro-differential equations, we need to find solutions to three systems of ordinary differential equations. Two systems describe the distribution of the field excited by one loop and the specified value of the excitation current in it. In the first system the loop is located at the section's input, and in the second, at the section's output. The third system does not depend on the loop parameters at all. It describes the distribution of the field excited by an electron beam in a section without loops.

Based on the developed analytical model, the computer code was developed for matching the loop couplers for the uniform accelerating sections of X-band. The calculation results were used to simulate the non-uniform section. Without additional matching, we obtained an input reflection coefficient of 8E-3.

Coupling elements always introduce a distortion in the field distribution and multi-pole components of the field can appear. So, for full modelling the couplers in an inhomogeneous accelerating section it is necessary to use 3D electromagnetic codes in the frequency domain [1]. In problems where axially symmetric fields are sufficient, approximate models such as the one described above can be useful as these fields weakly depend on the method of supplying high-frequency energy to and from the section (slits or loops)